\documentclass{amsart}
\usepackage{amssymb,latexsym}
\usepackage{graphicx}

\begin{document}

\title{Generalization of the Airy function and the operational methods}
\author{D. Babusci$^\dag$, G. Dattoli$^\ddag$, D. Sacchetti$^\diamond$} 

\address{$^\dag$ INFN - Laboratori Nazionali di Frascati, via E. Fermi 40, I-00044 Frascati.}
\email{danilo.babusci@lnf.infn.it}

\address{$^\ddag$ ENEA - Dipartimento Tecnologie Fisiche e Nuovi Materiali, Centro Ricerche Frascati\\
                 C. P. 65, I-00044 Frascati.}
 \email{giuseppe.dattoli@enea.it}
                 
\address{$^\diamond$ Dipartimento di Statistica Universit\`a 
                  ``Sapienza" di Roma, P.le A. Moro, 5, 00185 Roma.}
\email{dario.sacchetti@uniroma1.it}

\begin{abstract}
In this brief note the operatorial methods are applied to the study of the Airy function and 
its generalizations.
\end{abstract}

\maketitle

In this note we will discuss a method which can be usefully applied to the study of the Airy function. Before entering the 
details of the method we consider the following integral
\begin{equation}
\label{eq:Cint}
C(\alpha, \beta) \,=\, \int_0^\infty\,\mathrm{d}\xi\,\mathrm{e}^{\imath\,\xi^\alpha}\,\xi^\beta\;,
\end{equation}
which reduces to the ordinary Fresnel integral for $\alpha = 2$, $\beta = 0$. The use of standard analytical procedures 
allows to derive for it the following explicit expression in terms of the Gamma function
\begin{equation}
C(\alpha, \beta) \,=\,\frac{1}{\alpha}\,\Gamma\left(\frac{1 + \beta}{\alpha}\right)\,\exp\left\{\imath\,\frac{\pi}{2}\,
\frac{1 + \beta}{\alpha}\right\}\;,
\end{equation}
that will play a key role in the following.

Let us now consider the following integral transform 
\begin{equation}
\label{eq:Txalp}
T (x | \alpha) \,=\, \int_0^\infty\,\mathrm{d}\xi\,\mathrm{e}^{\imath\,\xi^\alpha}\,f(x\,\xi)
\end{equation}
which, on account of the operational identity \cite{Dat1}
\begin{equation}
\mathrm{e}^{\lambda\,x\,\partial_x}\,f(x) \,=\, f(\mathrm{e}^\lambda \,x)\;,
\end{equation}
can be written as \cite{Babu}
\begin{equation}
\label{eq:Ttrasf}
T (x | \alpha) \,=\,\int_0^\infty\,\mathrm{d}\xi\,\mathrm{e}^{\imath\,\xi^\alpha}\,\xi^{x\,\partial_x}\, f(x) 
\,=\, \hat{C} (\alpha, x\,\partial_x)\,f(x)\;,
\end{equation}
where we have assumed that the integral in eq. \eqref{eq:Cint} formally holds also when $ \beta$ is replaced by an operator 
(the integral itself is an operator). If the function $f(x)$ admits the expansion
\begin{equation}
f (x) \,=\, \sum_{n = 0}^\infty\,a_n\,x^n\;,
\end{equation}
we obtain (see ref. \cite{Babu})
\begin{equation}
\label{eq:Tsum}
T (x | \alpha) \,=\,\sum_{n = 0}^\infty\,a_n\,C(\alpha, n)\,x^n
\end{equation}
which provides an appropriate series expansion for the integral transform in eq. \eqref{eq:Txalp}.

The Airy function is defined through the expression \cite{Vall}
\begin{equation}
\mathrm{Ai} (x) \,=\,\frac{1}{2 \pi}\,\int_{-\infty}^\infty\,\mathrm{d}\xi\,\exp\left(\imath\,\frac{\xi^3}{3}\,+\,\imath\,x\,\xi\right)
\end{equation}
which is easily shown to satisfy the differential equation
\begin{equation}
y^{\prime\prime} \,-\,x\,y \,=\, 0\;.
\end{equation}

According to eqs. \eqref{eq:Ttrasf} and  \eqref{eq:Tsum} we can expand the Airy function as follows
\begin{eqnarray}
\mathrm{Ai} (x) \!\!&=&\!\! \frac{\sqrt[3]{3}}{\pi}\,\Re\left\{\int_0^\infty\,\mathrm{d}t\,\mathrm{e}^{\imath\,t^3}\,
\left(\sqrt[3]{3}\,t\right)^{x\,\partial_x}\,\mathrm{e}^{\imath\,x}\right\}\\
              &=&\!\! \frac{1}{\sqrt[3]{9}\,\pi}\,\sum_{n = 0}^\infty\,\frac{1}{n!}\Gamma\left(\frac{n + 1}{3}\right)\,
                       \cos\left(\frac{4n + 1}{6}\,\pi\right)\,(\sqrt[3]{3}\,x)^n\;. \nonumber
\end{eqnarray}
(For further comments on earlier derivation see ref. \cite{Vall}).

In the past, generalizations of the Airy function satisfying, for example, equations of the type
\begin{equation}
y^{\prime\prime} \,+\,c_n\,x^n\,y \,=\, 0\;.
\end{equation}
have been proposed by Watson \cite{Wats}. We consider first the example
\begin{equation}
\mathrm{Ai}_4 (x) \,=\,\int_0^\infty\,\mathrm{d}t\,\cos\left(t^4 \,+\, 2\,x\,t \,+\, 2\,x^2\right)
\end{equation}
which, on account of the previously outlined procedure, can be cast in the form
\begin{eqnarray}
\mathrm{Ai}_4 (x) \!\!&=&\!\! \Re\left\{\mathrm{e}^{\imath\,2\,x^2}\,\hat{C} (4, x\,\partial_x)\,\mathrm{e}^{\imath\,2\,x}\right\} \\
                   &=&\!\! \frac{1}{4}\,\sum_{n = 0}^\infty\,\frac{1}{n!}\Gamma\left(\frac{n + 1}{4}\right)\,
                           \left\{\cos(2\,x^2)\,\cos\left(\frac{5 n + 1}{8}\,\pi\right)\right. \nonumber \\
                   & &\qquad \qquad\qquad\qquad\quad \,-\, \left. \sin(2\,x^2)\,\cos\left(\frac{5 n + 1}{8}\,\pi\right)\right\}\,(2\,x)^n
                           \;. \nonumber
\end{eqnarray}

Another example is represented by the function defined by the following integral representation
\begin{equation}
P (x,y) \,=\,\int_0^\infty\,\mathrm{d}u\,\mathrm{e}^{\imath(u^4 \,+\, x\,u^2 \,+\,y\,u)}
\end{equation}
introduced by Pearcey (see \cite{Vall} and references therein) in the context of electromagnetic field theory. In this 
case, we obtain
\begin{eqnarray}
\label{eq:Pear}
P (x,y) \!\!&=&\!\! \hat{C} (4, 2\,x\,\partial_x \,+\, y\,\partial_y)\,\mathrm{e}^{\imath\,(x\,+\,y)}  \nonumber \\
                &=&\!\! \frac{\mathrm{e}^{\imath\,\pi/8}}{4}\,\sum_{n = 0}^\infty\,\mathrm{e}^{\imath\,3\,n\,\pi/4}\,x^n\,
                            \sum_{k = 0}^n\,\frac{1}{k!\,(n - k)!}\,\Gamma\left(\frac{2n - k + 1}{4}\right)\,
                            \left(\mathrm{e}^{-\imath\,\pi/8}\,\frac{y}{x}\right)^k\;. 
\end{eqnarray}
It is interesting to note that $P(x,y)$ satisfies a Schr\"{o}dinger-like equation
\begin{equation}
\imath\,\partial_x\,P (x,y) \,=\, \partial^2_y\,P (x,y)\;,
\end{equation}
and, therefore, we can write
\begin{equation}
\label{eq:PSchr}
P (x,y) \,=\, \mathrm{e}^{- \imath\,x\,\partial^2_y}\,P (0,y)\;.
\end{equation}
This result allows to write an alternative series expansion for $P(x,y)$. We first note that 
\begin{equation}
P (0,y) \,=\, \hat{C} (4, y\,\partial_y)\,\mathrm{e}^{\imath\,y} \,=\, \frac{\mathrm{e}^{\imath\,\pi/8}}{4}\,
\sum_{n = 0}^\infty\,\frac{1}{n!}\,\Gamma\left(\frac{n + 1}{4}\right)\,\left(\mathrm{e}^{\imath\,5\,\pi/8}\, y\right)^n\,.
\end{equation}
Moreover, from eq. \eqref{eq:PSchr} and the operational identity defining the generalized Hermite polynomials \cite{Dat2}
\begin{equation}
\mathrm{e}^{w\,\partial^2_z}\,z^n \,=\, H_n (z,w)\;, \qquad \qquad 
H_n (z,w) \,=\, n!\,\sum_{k = 0}^{[n/2]}\,\frac{1}{( n - 2k)!\,k!}\,z^{n - 2k}\,w^k\;,
\end{equation}
we, finally, get 
\begin{equation}
P (x,y) \,=\, \frac{\mathrm{e}^{\imath\,\pi/8}}{4}\,\sum_{n = 0}^\infty\,\frac{\mathrm{e}^{\imath\,5\,n\,\pi/8}}{n!}\,
\Gamma\left(\frac{n + 1}{4}\right)\,H_n(y, -\imath\,x).
\end{equation}

This brief note has been aimed at providing the possibility of treating Airy type integral in a unified way. A forthcoming, more 
extended, note will treat further relevant consequences.

\vspace{1.0cm}

\end{document}